\documentclass[a4paper,12pt]{article}
\usepackage{amssymb,amsmath}
\usepackage{hyperref}
\usepackage[dvipsnames]{xcolor}

\usepackage[utf8]{inputenc}
\usepackage[T2A]{fontenc}
\usepackage[english]{babel}
\usepackage[numbers]{natbib}
\usepackage{graphicx}
\hypersetup{linkcolor=Plum,citecolor=OliveGreen,urlcolor=NavyBlue,colorlinks=true}

\begin{document}
	\title{Hyperbolic and Elliptic Points Tracking Algorithm (HEPTA) in two-dimensional non-stationary velocity fields defined on a discrete grid}
	\author{\textsc{Udalov A.A., Uleysky M.Yu.} \\
		V.I. Il'ichev Pacific Oceanological Institute of the Russian \\Academy of Sciences , 43 Baltiyskaya St., Vladivostok, 
		690041, Russia 
	}
	\date{}
	\maketitle
	
	\begin{abstract}
	This article presents a new algorithm, the Hyperbolic and Elliptic Points Tracking Algorithm (HEPTA), designed for automated tracking of elliptic and hyperbolic stationary points in two-dimensional non-stationary velocity fields defined on a discrete grid. HEPTA analyzes the stability, bifurcations, and Lagrangian dynamics of stationary points. By leveraging bilinear interpolation, Jacobian matrix analysis, and trajectory tracking, the algorithm accurately identifies the locations of vortex centers (elliptic points) and strain zones (hyperbolic points). A methodology has been developed  to address bifurcation events and transitions across grid cell boundaries that occur during the evolution of stationary points in a discrete velocity field. The algorithm was tested on AVISO satellite altimetry data in the Kuroshio Current region, which is characterized by intense eddy formation. These data represent a two-dimensional discrete velocity field with a daily time step. The results show that HEPTA accurately identifies and tracks both cyclonic and anticyclonic eddies, even under conditions of rapid eddy drift and complex hydrodynamic conditions. This study provides a reliable and efficient tool for analyzing the dynamics of mesoscale formations, which may be useful in oceanographic research, climate modeling, and operational oceanography.
		
	\end{abstract}
	
	\section{Introduction}
	The use of satellite altimetry has facilitated the transition from local studies of the World Ocean to global-scale research. This approach enables continuous, high-accuracy monitoring of sea surface height through radar altimeters onboard Earth-orbiting satellites. The development of altimetric technologies began in the 1970s, with significant advancements achieved via the TOPEX/Poseidon program launched in 1992 \cite{Robinson2010}. For decades, altimetry data have been instrumental in studying coherent ocean structures, which critically influence water mass transport and mixing processes \cite{Prants2006, Dijkstra2000}.

	In oceanography, important coherent structures include eddies and hyperbolic Lagrangian coherent structures (LCSs) \cite{Haller2012}. The latter act as transport barriers, separating water masses of distinct origins and properties. These barriers are crucial for studying both pollution dispersion \cite{Fukushima2011, Chazhma2022} and fisheries dynamics \cite{Kulik2022}.

	For slowly varying velocity fields, the centers of eddies and hyperbolic LCSs can be approximated as elliptic and hyperbolic points of the instantaneous velocity field.  Eddies are coherent rotating structures capable of transporting physico-chemical properties (e.g., heat, salt, and biochemical tracers) across oceanic regions \cite{Udalov2023}. Over the past decades, several Lagrangian methods have been developed for identifying oceanic eddies from satellite data, including those based on: the finite-time Lyapunov exponent \cite{Shadden2005}, the finite-scale Lyapunov exponent \cite{Ovidio2009}, and the Lagrangian-averaged vorticity deviation \cite{Haller2016}, as well as a method for determining the eddy center via elliptic point in the altimetric velocity fields \cite{Budyansky2024}.
	
	Hyperbolic LCSs are also observed in the ocean, manifesting as zones of maximum strain (compression and stretching) on two-dimensional surfaces, such as the sea surface \cite{Haller2012}. In studies of marine debris dispersion, regions near hyperbolic LCSs help pinpoint accumulation hotspots \cite{Kunz2024, Serra2020}. Conversely, these structures can also enhance pollutant dispersion through stretching mechanisms \cite{Fukushima2011, Duran2021, Prants2023}.
	
	By addressing the challenge of detecting and tracking elliptic/hyperbolic points in slowly varying velocity fields, we enable efficient monitoring of eddies and hyperbolic LCSs with minimal computational overhead.

	In this work, we present HEPTA (Hyperbolic and Elliptic Points Tracking Algorithm), a novel method for tracking the spatio-temporal evolution and bifurcations of elliptic/hyperbolic points in discrete velocity fields. HEPTA effectively resolves the dynamics of both point types, offering new insights into their interactions. Using this approach, we can investigate the influence of the two types of points on each other, thereby capturing the interaction of eddy and hyperbolic LCSs.

	This document is organized as follows. Section 2 details the methodology for identifying stationary points in a discrete two-dimensional velocity field, followed by their classification based on stability analysis. Section 3 is dedicated to a detailed description of the main stages of the HEPTA algorithm, which is designed for tracking the evolution of elliptic and hyperbolic points in two-dimensional non-stationary velocity fields. Section 4 presents the results of applying the HEPTA algorithm to track a large cyclonic eddy based on AVISO altimetric velocity field data. Section 5 contains a discussion of the algorithm, including its advantages, limitations, and directions for further development. In conclusion, the main results of the study are summarized.

	\section{Determination and stability analysis of instantaneous stationary points in a discrete two-dimensional velocity field}
	
	Elliptic and hyperbolic Lagrangian coherent structures play a key role in analyzing the dynamics of stationary fluid or gas flows. These structures are associated with points where the velocity vector $\vec{v} = (u(x,y,t), v(x,y,t))$ becomes zero, allowing us to determine the nature of the flow. Specifically, they can mark transition zones between stable and unstable regimes, and also act as centers of vortex structures (elliptic points) or saddle points (hyperbolic points), forming the overall flow pattern. For non-stationary, slowly evolving flows, stationary points also retain their significance, as they are located near the centers of eddies and intersection points of "Lyapunov ridges" in forward and backward time.
	
	This chapter presents a method for determining the position and analyzing the type of stability of stationary points at a fixed time instant for a two-dimensional discrete velocity field defined on a regular rectangular grid. Determining stationary points is a necessary preliminary step for constructing algorithms to track them over time, which will allow us to study the evolution of LCSs.
	
	It is assumed that the velocity field is given at the nodes of a two-dimensional regular grid. At each node, the values of the velocity vectors are known: $\vec{v}_{00} = (u_{00},v_{00})$, $\vec{v}_{01} = (u_{01},v_{01})$, $\vec{v}_{10} = (u_{10},v_{10})$, $\vec{v}_{11} = (u_{11},v_{11})$. For ease of computation, the coordinates $x$ and $y$ are normalized and vary from 0 to 1. To approximate the behavior of the flow inside the cell, bilinear interpolation is used, which leads to the following system of equations:
	\begin{equation} \label{1}
		\begin{pmatrix}
			\dot{x} \\
			\dot{y}
		\end{pmatrix} =
		\begin{pmatrix}
			a_x \\
			a_y
		\end{pmatrix}xy +
		\begin{pmatrix}
			b_x \\
			b_y
		\end{pmatrix}x +
		\begin{pmatrix}
			c_x \\
			c_y
		\end{pmatrix}y +
		\begin{pmatrix}
			d_x \\
			d_y
		\end{pmatrix},
	\end{equation}
	where the coefficients $a_x, a_y, b_x, b_y, c_x, c_y, d_x, d_y$ are expressed in terms of the velocity values at the grid nodes:
	\begin{displaymath}
		\begin{gathered}
			\begin{pmatrix}
				a_x \\
				a_y
			\end{pmatrix} = \vec{v}_{11}+\vec{v}_{00}-\vec{v}_{01}-\vec{v}_{10},\quad
			\begin{pmatrix}
				b_x \\
				b_y
			\end{pmatrix}= \vec{v}_{10} - \vec{v}_{00},\\
			\begin{pmatrix}
				c_x \\
				c_y
			\end{pmatrix}=\vec{v}_{01} - \vec{v}_{00},\quad
			\begin{pmatrix}
				d_x \\
				d_y
			\end{pmatrix}=\vec{v}_{00}.
		\end{gathered}
	\end{displaymath}
	
	To determine the coordinates of the stationary points, it is necessary to equate to zero the velocity projections given by equation  \eqref{1}.
	\begin{equation}
		\begin{aligned}
			a_xxy+b_xx+c_xy+d_x&=0,\\
			a_yxy+b_yx+c_yy+d_y&=0.
		\end{aligned}
		\label{1a}
	\end{equation}
	To express the variable $y$ in terms of $x$, we eliminate the nonlinear term $xy$ from the system~\eqref{1a}. To do this, we multiply the first equation by $a_y$, and the second by $a_x$. If we subtract the second equation from the first, we obtain the following expression for $y$:
	
	\begin{equation}
		y = \dfrac{(a_x b_y - a_y b_x)x + (a_x d_y - a_y d_x)}{a_y c_x - a_x c_y}.
		\label{1b}
	\end{equation}
	If $a_y c_x - a_x c_y = 0$, the expression for $y$ becomes undefined. Therefore, we will only consider cases where $a_y c_x - a_x c_y \neq 0$. Substituting expression~\eqref{1b}  into the first equation of the system~\eqref{1a},we obtain a quadratic polynomial with respect to the variable  $x$:
	\begin{equation*}
		\alpha x^2 + \beta x + \gamma=0,
	\end{equation*}
	
	where $\alpha = a_x^2 b_y - a_x a_y b_x $, $\beta=a_x^2 d_y - a_x a_y d_x - a_x b_x c_y + a_y b_x c_x + a_x b_y c_x - a_y b_x c_x $,  $\gamma = a_y c_x d_x - a_x c_y d_x + a_x c_x d_y - a_y c_x d_x$. Solving this quadratic equation allows us to find the abscissas $x_0$ of the stationary points:
	\begin{equation}\label{2_x}
		x_0 = \dfrac{1}{2\alpha}(-\beta\pm\sqrt{\beta^2-4\alpha\gamma}).
	\end{equation}
	In the case where $\alpha = 0$, the value of the abscissa is found from the linear equation and is equal to $x_0 = -\gamma/\beta$ (provided that $\beta \neq 0$). The values of the ordinates $y_0$ are found by substituting the found values of $x_0$ into equation~\eqref{1b}.
	
	Since the interpolation function~\eqref{1} only makes sense within the boundaries of the cell, additional conditions $0\leqslant x_0\leqslant 1$ and $0\leqslant y_0\leqslant 1$are imposed on the stationary points. Thus, there can be from zero to two stationary points inside each cell.
	
	To determine the type of stability of the found stationary points 	 $(x_0,y_0)$, we linearize the system~\eqref{1} in the vicinity of the stationary position:
	\begin{equation}\label{3}
		\begin{pmatrix}
			\Delta\dot{x} \\
			\Delta\dot{y}
		\end{pmatrix}= J(x_0,y_0)
		\begin{pmatrix}
			\Delta x \\
			\Delta y
		\end{pmatrix},
	\end{equation}
	where
	\begin{equation*}
		J(x_0,y_0) =
		\begin{pmatrix}
			a_x y_0 +b_x &a_x x_0 + c_x \\
			a_y y_0 + b_y & a_y x_0+ c_y
		\end{pmatrix}
	\end{equation*}
	is the Jacobian matrix of the system~\eqref{1} at the stationary point $(x_0,y_0)$. The type of stability of the stationary point is determined by the eigenvalues $\lambda_1$ and $\lambda_2$  of the matrix $J$. If both eigenvalues are real ($\lambda_1$, $\lambda_2 \in \mathbb{R}$) and have different signs  ($\lambda_1\lambda_2 < 0$), the stationary point is a saddle with a pair of stable and unstable manifolds. In the case of real eigenvalues of the same sign, we are dealing with a stable or unstable node. If the eigenvalues are a complex conjugate pair, the stationary point will be a stable or unstable focus, being the center of a vortex structure.
	
	In oceanographic problems, the direction of flow rotation near an elliptic stationary point is also of interest. This direction can be determined, for example, by the sign of $\Delta\dot y$ at the point ($\Delta x=1$, $\Delta y=0$), i.e., by the sign of the expression $a_yy_0+b_y$. If $a_yy_0+b_y>0$, the eddy is cyclonic (counterclockwise), otherwise it is anticyclonic (clockwise).
	
	\section{Construction of an algorithm for tracking the evolution of elliptic and hyperbolic LCSs}
	
	In the previous chapter, a method was presented for determining the position and classifying stationary points in a two-dimensional discrete velocity field defined on a regular grid at a fixed moment in time. However, for a comprehensive analysis of the velocity field dynamics, it is necessary to trace the evolution of these stationary points over time, tracking their trajectories and accounting for possible changes in their characteristics. The tracking algorithm must consider a number of key aspects, including bifurcations, transitions across cell boundaries, and interactions with the boundaries of the study area.

	The first key aspect that must be considered when constructing an algorithm for tracking stationary points is the handling of bifurcations. In system \eqref{1}, only the simplest bifurcations of the birth/destruction of a pair of points are possible. One of the points will always be a saddle, the second either a node or a focus. The algorithm must be able to identify such events, record the fact of bifurcation, and correctly link stationary points before and after bifurcation, if possible. Incorrect handling of bifurcations can lead to the termination of tracking individual stationary points and an incorrect interpretation of the flow dynamics.

	The second key aspect is the correct tracking of the movements of stationary points between spatial cells of the grid. Since the velocity field is discretized on a regular grid, stationary points can migrate from one cell to another in the process of the evolution of the velocity field. The algorithm must continuously track the position of each stationary point and determine when a transition across the cell boundary occurs. When such a transition is detected, it is necessary to transfer information about the tracked point to the neighboring cell, thereby ensuring the continuity of the trajectory of the stationary point in time. Incorrect tracking of transitions between cells can lead to a break in the trajectory and termination of tracking of the point.

	Finally, the algorithm must take into account the potential interactions of stationary points both with the boundaries of the study area and with the coastline, if any. This includes scenarios in which a stationary point may disappear when crossing the boundary, as well as the possibility of new points appearing inside the region.

	Combining the methods for determining and classifying stationary points presented in the previous chapter with a mechanism for accounting for bifurcations, transitions across cell boundaries, and interactions with the boundaries of the region allows us to build a stable and reliable algorithm for tracking elliptic and hyperbolic Lagrangian coherent structures in time. Such an algorithm will allow for a detailed analysis of the evolution of LCSs and their impact on the dynamics of the system under study.

	\subsection{Stationary point birth and death events}
	
	Consider a system of equations describing the behavior of the flow in the case where the velocity vector field explicitly depends on time. Since the velocity field is given discretely in time, events occurring in the interval between two consecutive time moments, $t\in [t_1,t_2]$, where $t_1$ and $t_2$ are two closest time moments for which velocity field data is available, are of particular interest. Trilinear interpolation is applied to approximate the values of the vector field at an arbitrary point in space and at any time within this interval.
	
	Trilinear interpolation requires knowledge of the velocity vector values at eight nodes: four nodes defining the grid cell at times $t_1$ and $t_2$, respectively. Based on this interpolation, we obtain the following system of equations describing the temporal evolution of the coordinates:
	\begin{eqnarray}\label{t}
		\begin{pmatrix}
			\dot{x} \\
			\dot{y}
		\end{pmatrix} =
		\begin{pmatrix}
			a_x + \tau (a'_x - a_x) \\
			a_y + \tau (a'_y - a_y)
		\end{pmatrix}xy +
		\begin{pmatrix}
			b_x + \tau (b'_x - b_x) \\
			b_y + \tau (b'_y - b_y)
		\end{pmatrix}x + \nonumber \\[5pt]
		+
		\begin{pmatrix}
			c_x + \tau (c'_x - c_x) \\
			c_y + \tau (c'_y - c_y)
		\end{pmatrix}y +
		\begin{pmatrix}
			d_x + \tau (d'_x - d_x) \\
			d_y + \tau (d'_y - d_y)
		\end{pmatrix},
	\end{eqnarray}
	where the parameters without primes are defined at time $t_1$, with primes at time $t_2$, and $\tau = (t - t_1)/(t_2 - t_1)$ is the normalized time, varying in the range from 0 to 1. The expressions for calculating the parameters $a,b,c,d$ are similar to those described in equation~\eqref{1}.
	
	By analogy with the stationary case considered in section two, we can determine the positions of stationary points at time $\tau$, considering $\tau$ as a parameter varying in the interval $[0,1]$.  Stationary points are defined as solutions to the system of equations~\eqref{t}, where the velocity projections are zero. We represent the system~\eqref{t} as
	\begin{equation}
		\begin{aligned}
			\tilde{a}_xxy+\tilde{b}_xx+\tilde{c}_xy+\tilde{d}_x&=0,\\
			\tilde{a}_yxy+\tilde{b}_yx+\tilde{c}_yy+\tilde{d}_y&=0.
		\end{aligned}
		\label{ta}
	\end{equation}
	where, for simplicity, we have switched to the coefficients $\tilde{a}_x = a_x + \tau (a'_x - a_x)$, $\tilde{b}_x = b_x + \tau (b'_x - b_x)$, $\tilde{c}_x = c_x + \tau (c'_x - c_x)$, $\tilde{d}_x = d_x + \tau (d'_x - d_x)$ and similarly for $\tilde{a}_y, \tilde{b}_y, \tilde{c}_y, \tilde{d}_y$.  Eliminating the nonlinear term $xy$(see section 2), we obtain a parametric dependence of the coordinates of the stationary points on $\tau$:
	\begin{equation}\label{xyt_12}
		\begin{cases}
			x_0(\tau) = \dfrac{1}{2\alpha(\tau)}(-\beta(\tau)\pm\sqrt{\beta^2(\tau)-4\alpha(\tau)\gamma(\tau)} ),\\[12pt]
			y_0(\tau)= \dfrac{(\tilde{a}_x \tilde{b}_y - \tilde{a}_y \tilde{b}_x)x_0(\tau) + (\tilde{a}_x \tilde{d}_y - \tilde{a}_y \tilde{d}_x)}{\tilde{a}_y \tilde{c}_x - \tilde{a}_x \tilde{c}_y}
		\end{cases}
	\end{equation}
	where $\alpha(\tau) = \tilde{a}_x^2 \tilde{b}_y - \tilde{a}_x \tilde{a}_y \tilde{b}_x $, $\beta(\tau)=\tilde{a}_x^2 \tilde{d}_y - \tilde{a}_x \tilde{a}_y \tilde{d}_x - \tilde{a}_x \tilde{b}_x \tilde{c}_y + \tilde{a}_y \tilde{b}_x \tilde{c}_x + \tilde{a}_x \tilde{b}_y \tilde{c}_x - \tilde{a}_y \tilde{b}_x \tilde{c}_x $,  $\gamma(\tau) = \tilde{a}_y \tilde{c}_x \tilde{d}_x - \tilde{a}_x \tilde{c}_y \tilde{d}_x + \tilde{a}_x \tilde{c}_x \tilde{d}_y - \tilde{a}_y \tilde{c}_x \tilde{d}_x$. Similar to the stationary case, it is necessary to discard the degenerate case when $\tilde{a}_y \tilde{c}_x - \tilde{a}_x \tilde{c}_y = 0$. When $\alpha = 0$, we should move on to considering the linear case for finding the abscissa $x_0(\tau)$. At $\tau = 0$ the system will move to the case considered in section 2.
	
	We are interested in events related to bifurcations, that is, changes in the number of stationary points in the cell during the time interval $\tau \in [0,1]$. In particular, we consider the cases when the number of stationary points changes from 0 to 2 (birth of a pair) or from 2 to 0 (disappearance of a pair). To determine the moments of bifurcations, it is necessary to find the values of time $\tau$, at which the discriminant $D(\tau) = \beta^2(\tau)-4\alpha(\tau)\gamma(\tau)$ becomes zero.
	
	The discriminant $D(\tau)$, which determines the moments of bifurcations of stationary points, is formally a polynomial with respect to the parameter $\tau$. Each of the coefficients $\alpha(\tau),\beta(\tau),\gamma(\tau)$  depends on $\tau$ through linear combinations of the original parameters of the system. Since the coefficients $\tilde{a}_x, \tilde{b}_x, \tilde{c}_x, \tilde{d}_x$ (and their analogs for $y$) depend linearly on $\tau$,  their products in the expressions for $\alpha(\tau),\beta(\tau),\gamma(\tau)$ generate polynomial terms of higher degrees. For example, $\alpha(\tau)$ and $\beta(\tau)$ turn out to be polynomials of the third degree, and $\gamma(\tau)$ is of the fourth degree.
	
	At first glance, $D(\tau)$ must have degree $3 \times 2 + 4\times 3 = 7$ (since $\beta^2(\tau)$ is the square of a third-degree polynomial, and $\alpha(\tau)\gamma(\tau)$ is the product of third- and fourth-degree polynomials). However, due to the symmetry of the system and the algebraic properties of the coefficients, the higher-order terms mutually cancel out. This is because the terms responsible for the higher powers of $\tau$ in $\beta^2(\tau)$ and $\alpha(\tau)\gamma(\tau)$, turn out to be proportional to each other with opposite signs. As a result, the maximum degree of  $D(\tau)$ decreases to four.
	
	Since $D(\tau)$ is a 4th-degree polynomial with respect to time, the maximum number of bifurcations per interval $[0,1]$ does not exceed four (according to the fundamental theorem of algebra). However, it is important to take into account that the polynomial $D(\tau)$ may have complex or multiple roots, which affects the nature of the bifurcation. Two scenarios are possible in the system: the birth of a pair of stationary points (from 0 to 2), when $D(\tau)$ changes from negative to positive values, generating two new stationary solutions (hyperbolic and elliptic); and the disappearance of a pair of stationary points (from 2 to 0), when $D(\tau)$ changes from positive to negative values, leading to the merging and disappearance of two existing solutions.

	\subsection{Dynamics of stationary points and their transition across cell boundaries}
	
	\subsubsection{Algorithm for finding stationary points on boundaries}
	
	The regular grid on which the velocity field is defined divides the region into discrete cells. Stationary points found in each cell are not "tied" to it, and their position may change over time depending on the evolution of the velocity field. When the system parameters change (for example, the velocity at the cell boundaries), the stationary point may undergo the following changes: remain inside the cell, changing its position; move to the boundary; leave the cell, moving to the neighboring one.

	To correctly track the trajectories of stationary points, it is necessary to analyze the behavior of the solutions of system~\eqref{t} on all four boundaries of the cell: $ x = 0$, $x = 1$, $y = 0$, $y =1$.  Each boundary defines an additional condition, reducing the system to an equation with respect to one spatial parameter and time $\tau$. The method for finding the time moments when the stationary point will be on the boundary is similar to that described in section 2. We get rid of the nonlinear term by multiplying the two equations by the corresponding coefficients and then subtracting one equation from the other. The subsequent expression of one spatial variable through time $\tau$ leads to a quadratic equation for time:
	\begin{equation*}
		\alpha \tau^2 + \beta \tau + \gamma = 0,
	\end{equation*}
	where the values of $\alpha,\beta,\gamma$ will have different forms, depending on the boundary conditions. Further, in the derivation, we will replace all differences of coefficients $(a'_x - a_x)$ with the notation $\Delta a_x$, similarly for all other parameters $x$ and $y$.
	
	On the left boundary ($x=0$), the system of equations~\eqref{t} takes the form:
	\begin{equation}
		\begin{cases}
			\Delta c_x\tau y+c_xy+\Delta d_x\tau+d_x = 0, \\
			\Delta c_y\tau y+c_yy+\Delta d_y\tau+d_y = 0.
		\end{cases}
	\end{equation}
	where $\Delta c_x = (c'_x - c_x),\Delta d_x = (d'_x - d_x)$, similarly for the parameters $\Delta c_y ,\Delta d_y$. Solving this system allows us to find the possible values of $y_0$ and time $\tau_0$ at which the stationary point is on the boundary $x=0$.
	\begin{equation}
		\begin{cases}
			\tau_0 = \dfrac{1}{2\alpha}(-\beta \pm \sqrt{\beta^2 - 4\alpha\gamma}), \\[12pt]
			y_0 = \dfrac{ (\Delta d_x \Delta c_y - \Delta d_y \Delta c_x) \tau_0 + (d_x \Delta c_y - d_y \Delta c_x) }
			{ c_y \Delta c_x - c_x \Delta c_y },
		\end{cases}
	\end{equation}
	where $\alpha= \Delta c_x( d_x\Delta c_y - d_y\Delta c_x)$, $\beta = \Delta c_x(d_x\Delta c_y - d_y\Delta c_x)+c_x(\Delta d_x\Delta c_y -\Delta d_y\Delta c_x)+\Delta d_x(\Delta c_x c_y -\Delta c_y c_x)$, $\gamma= d_x(\Delta c_x c_y -\Delta c_y c_x)$. If the found value $\tau_0$ lies in the interval $[0,1]$, this means that the stationary point crosses the boundary $x=0$ at the corresponding time moment. The absence of real solutions (negative discriminant) indicates that the stationary point does not reach this boundary in the considered time interval $[t_1, t_2]$. It is also necessary to take into account that at $\alpha=0$ it is required to consider a linear equation for finding the crossing time. We exclude from consideration the case when $\Delta c_x c_y -\Delta c_y c_x =0$.
	
	Similarly, on the right boundary ($x=1$),  the system of equations also simplifies, since the coordinate  $x$  becomes fixed and equal to 1. The solution in this case includes more complex expressions, due to the presence of both linear and nonlinear terms in equation~\eqref{t}:
	\begin{equation}
		\begin{cases}
			(\Delta c_x + \Delta a_x)\tau y+(\Delta d_x + \Delta b_x)\tau+(c_x+a_x)y+d_x+b_x = 0, \\
			(\Delta c_y + \Delta a_y)\tau y+(\Delta d_y + \Delta b_y)\tau+(c_y+a_y)y+d_y+b_y = 0.
		\end{cases}
	\end{equation}
	
	Solving this system allows us to find the possible values of $y_0$ and time $\tau_0$ at which the stationary point is on the boundary $x = 1$.
	\begin{equation}
		\begin{cases}
			\tau_0 = \dfrac{1}{2\alpha} \left( -\beta \pm \sqrt{\beta^2 - 4\alpha\gamma} \right), \\[12pt]
			y_0 = \dfrac{ \left( (\Delta b_x - \Delta d_x)(\Delta a_y - \Delta c_y) - (\Delta b_y - \Delta d_y)(\Delta a_x - \Delta c_x) \right) \tau_0 }
			{ (a_y + c_y)(\Delta a_x + \Delta c_x) - (a_x + c_x)(\Delta a_y + \Delta c_y) }
			+ \\[10pt]
			\quad + \dfrac{ (b_x + d_x)(\Delta a_y + \Delta c_y) - (b_y + d_y)(\Delta a_x + \Delta c_x) }
			{ (a_y + c_y)(\Delta a_x + \Delta c_x) - (a_x + c_x)(\Delta a_y + \Delta c_y) },
		\end{cases}
	\end{equation}
	where $\alpha=(\Delta a_x + \Delta c_x)((\Delta b_x + \Delta d_x)(\Delta a_y + \Delta c_y)-(\Delta b_y+\Delta d_y)(\Delta a_x+\Delta c_x))$, $\beta=(\Delta a_x+\Delta c_x)((b_x+d_x)(\Delta a_y+\Delta c_y)-(b_y+d_y)(\Delta a_x+\Delta c_x))+(\Delta b_x+\Delta d_x)((a_y+c_y)(\Delta a_x+\Delta c_x)-(a_x+c_x)(\Delta a_y+\Delta c_y))+(a_x+c_x)((\Delta b_x+\Delta d_x)(\Delta a_y+\Delta c_y)-(\Delta b_y+\Delta d_y)(\Delta a_x+\Delta c_x))$ and $\gamma=(b_x+d_x)((a_y+c_y)(\Delta a_x+\Delta c_x)-(a_x+c_x)(\Delta a_y+\Delta c_y))$. For the analysis of transitions across the right boundary, it is also necessary to take into account the cases when $\alpha=0$ and $(a_y+c_y)(\Delta a_x + \Delta c_x) - (a_x+c_x)(\Delta a_y + \Delta c_y)=0$. The complexity of these equations reflects the influence of all velocity components (both linear and nonlinear, arising from bilinear interpolation) on the movement of the stationary point. This emphasizes the importance of taking into account the relationships between the parameters $a,b,c,d$, which encode information about the velocity field at the grid nodes.

	The equations for the horizontal boundaries $y = 0$, $ y = 1$ are symmetric to the equations for the vertical boundaries  $x = 0$, $x = 1$ with the replacement of parameters $b \leftrightarrow c$. This symmetry corresponds to the permutation of the contributions of the longitudinal and transverse velocity components to the system of equations. As an example for comparison, consider also the boundary condition when $y = 0$. A similar approach is applied for the horizontal boundaries of the cell ($y = 0$, $y = 1$). For example, at $y = 0$, substituting this condition into equation~\eqref{t} and solving for the coordinate $x_0$ and time $\tau_0$, we obtain:
	\begin{equation}
		\begin{cases}
			\tau_0 = \dfrac{1}{2\alpha}(-\beta \pm \sqrt{\beta^2 - 4\alpha\gamma}), \\[12pt]
			x_0 = \dfrac{(\Delta d_x\Delta b_y -\Delta d_y\Delta b_x)\tau_0 + d_x\Delta b_y - d_y\Delta b_x}{(\Delta b_x b_y -\Delta b_y b_x)},
		\end{cases}
	\end{equation}
	where $\alpha= \Delta b_x( d_x\Delta b_y - d_y\Delta b_x)$, $\beta = \Delta b_x(d_x\Delta b_y - d_y\Delta b_x)+b_x(\Delta d_x\Delta b_y -\Delta d_y\Delta b_x)+\Delta d_x(\Delta b_x b_y -\Delta b_y b_x)$, $\gamma= d_x(\Delta b_x b_y -\Delta b_y b_x)$. These expressions allow us to find the coordinate $x_0$ and the time moment $\tau_0$ at which the stationary point crosses the boundary. It is necessary to take into account the cases when $\alpha=0$ and $(\Delta b_x b_y -\Delta b_y b_x)=0$.

	\subsubsection{Direction of movement of stationary points on the boundary}
	
	To predict the further trajectory of a stationary point after crossing the cell boundary, it is necessary to determine the direction of its movement at the moment of crossing. This is achieved by linearizing the system of equations~\eqref{t} in the vicinity of the boundary crossing point ($x_0,y_0,\tau_0$). The Taylor series expansion allows expressing small increments of coordinates $\Delta x, \Delta y$ through the increment of time $\Delta \tau$. Since the stationary point is on the boundary, the change in coordinates must satisfy the following system of equations:
	\begin{equation}
		\begin{cases}
			\frac{\partial f_1}{\partial x}\Delta x+\frac{\partial f_1}{\partial y}\Delta y+\frac{\partial f_1}{\partial t}\Delta \tau = 0, \\
			\frac{\partial f_2}{\partial x}\Delta x+\frac{\partial f_2}{\partial y}\Delta y+\frac{\partial f_2}{\partial t}\Delta \tau = 0,
		\end{cases}
	\end{equation}
	where $f_1 = \dot{x}, f_2 = \dot{y}$ are the velocity components defined by equation~\eqref{t}, and the partial derivatives are calculated at the point ($x_0,y_0,\tau_0$).
	
	Solving this system of equations with respect to $\Delta x$ and $\Delta y$, we get:
	\begin{equation}
		\Delta x =\frac{\frac{\partial f_1}{\partial \tau}\frac{\partial f_2}{\partial y}-\frac{\partial f_1}{\partial y}\frac{\partial f_2}{\partial \tau}}{\operatorname{det}(J)}\Delta \tau
	\end{equation}
	\begin{equation}
		\Delta y =\frac{\frac{\partial f_1}{\partial \tau}\frac{\partial f_2}{\partial x}-\frac{\partial f_1}{\partial x}\frac{\partial f_2}{\partial \tau}}{\operatorname{det}(J)}\Delta \tau
	\end{equation}
	where $\operatorname{det}(J) = \dfrac{\partial f_1}{\partial y}\dfrac{\partial f_2}{\partial x}-\dfrac{\partial f_1}{\partial x}\dfrac{\partial f_2}{\partial y}$ is the determinant of the Jacobian matrix of the system, calculated at the point ($x_0,y_0,\tau_0$). It is important to note that if $\operatorname{det}(J)=0$, then further analysis is needed using higher-order terms in the Taylor expansion. The sign of $\Delta x, \Delta y$ determines the direction of the point's movement. For example, $\Delta x > 0$ indicates a shift to the right, $\Delta y < 0$ indicates a shift downwards, given $\Delta \tau >0$
	
	\subsection{Algorithm for analyzing the dynamics of stationary points}
	
	After presenting methods for determining the position of stationary points, analyzing their stability, and accounting for bifurcations (birth/death) of these points, we proceed to describe an algorithm for analyzing the dynamics of stationary points in a two-dimensional discrete velocity field. The algorithm is based on bilinear interpolation, which allows finding stationary points inside each cell. This algorithm, combining all the previously considered stages into a single system, provides automated analysis of the dynamics of stationary points. The flowchart of the algorithm is shown in Fig. \ref{fig:diag}, and a detailed description of each step is given below.

	\begin{figure}
		\centering
		\includegraphics[width=0.8\linewidth]{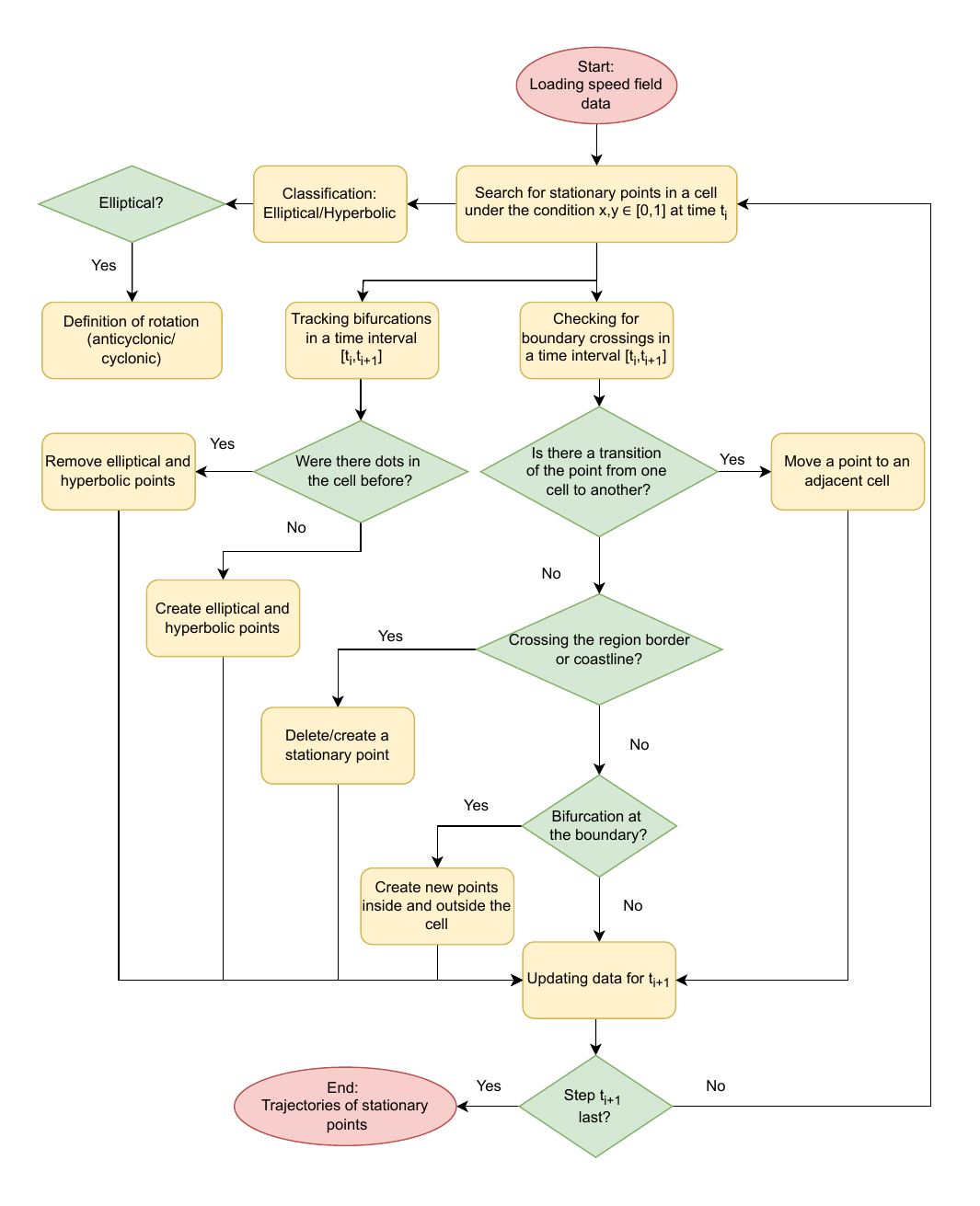}
		\caption{Flowchart of the HEPTA algorithm. The diagram shows rectangles (basic calculations) and rhombuses (conditional operators). The beginning and end of the algorithm are marked with red ellipses}
		\label{fig:diag}
	\end{figure}
	
	The first step of the algorithm is to determine the stationary points at each time step. For each time moment $t_i$  a search for stationary points is performed in each cell of the grid using bilinear interpolation (as described in section 2, by finding points where the interpolated velocity components become zero). After that, points that go beyond the cell boundaries are discarded, that is, whose coordinates do not satisfy the condition $x,y\in[0,1]$. Then, the stability of the found stationary points is classified. For each point, the Jacobian matrix $J$  is calculated and its eigenvalues are determined, based on which the type of stability of the stationary point is classified (elliptic or hyperbolic). For elliptic stationary points, the direction of rotation (cyclonic or anticyclonic) is also determined.
	
	The next stage includes tracking bifurcations and transitions across cell boundaries. On the time interval $[t_i,t_{i+1}]$ bifurcation events (birth or death of points) are identified by analyzing the zeros of the discriminant $D(t)$. Simultaneously, boundary crossings ($x=0,1$; $y=0,1$) are recorded with a check of the direction of movement by calculating the increments $\Delta x$ and $\Delta y$.
	
	When crossing a boundary, the following situations are possible: if the point is recorded in both neighboring cells, it is transferred. Otherwise, two options are considered: either the point crossed the boundary of the region (or the coastline), or a bifurcation occurred on the cell boundary. In the latter case, two new stationary points appear simultaneously, one of which begins to move inside the cell, and the other~--- outside. The algorithm must correctly identify all these events~--- both the transitions of stationary points between cells and bifurcations on the boundary, manifested as the crossing of cell boundaries by points.
	
	Finally, the trajectories of the stationary points are tracked. Their position is updated taking into account the bifurcations and transitions across cell boundaries that occurred at time $t_i$. This may include creating new trajectories for the appeared points, as well as merging or terminating existing ones. Then, the predicted coordinates of the stationary points are compared with the data at the next time moment $t_{i+1}$, which allows constructing trajectories of their movement.

	\section{Results}
	This section discusses the application of the HEPTA algorithm for automated identification and tracking of mesoscale eddy centers. The initial data are altimetric measurements of the AVISO/CMEMS velocity field, available on the portal \url{aviso.altimetry.fr}. These data are characterized by a spatial resolution of $0.25^\circ\times 0.25^\circ$ and daily temporal discretization, which makes it possible to reliably detect eddies with a diameter of 50~km or more \cite{Udalov2023,Chelton2011}.
	
	The eastern Pacific Ocean, located within the influence zone of the meandering Kuroshio Jet Stream, was selected as the study region. This choice stems from the area's active eddy formation processes. The dynamic instability of the Kuroshio Current, interactions of fast flows with bottom topography, and lateral branches of subtropical currents create favorable conditions for velocity shear zones. These zones, in turn, serve as sources for mesoscale eddy formation. Additionally, seasonal variability in wind forcing and baroclinic effects contribute to generating both anticyclonic and cyclonic structures.
	
\begin{figure}
		\centering
		\includegraphics[width=1\linewidth]{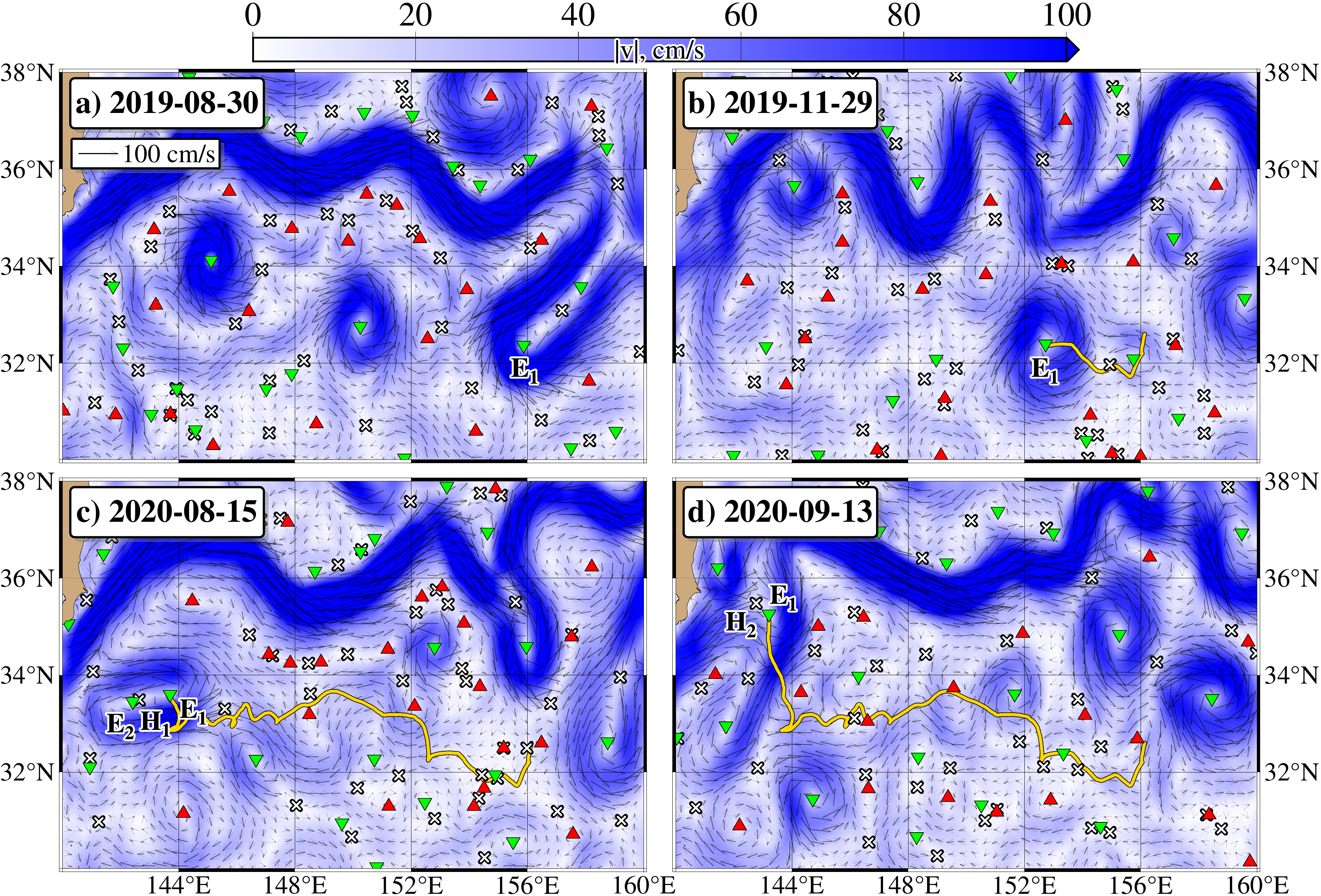}
		\caption{The figure shows the key stages of the evolution of the cyclonic eddy $E_1$. Panel \textbf{a} corresponds to the moment of the eddy's origin. Panel \textbf{b} reflects its position on November 29, 2019. Panel \textbf{c} shows the phase of separation of the eddy $E_1$: as a result of dynamic instability, a secondary cyclonic eddy $E_2$ and a hyperbolic point $H_1$ are formed. Panel \textbf{d} demonstrates the last stage of the existence of $E_1$ before its disappearance, when the eddy interacts with the hyperbolic point $H_2$. The arrows show the velocity field reconstructed from AVISO satellite altimetry data. The color background corresponds to the values of the length of the velocity vector. The gray line illustrates the trajectory of the center of the eddy $E_1$ from the moment of its origin to the current stage of evolution. Red triangles~--- anticyclonic eddies, green~--- cyclonic eddies, crosses~--- hyperbolic points}
		\label{fig:2}
	\end{figure}

	As an example of the analysis of the dynamics of mesoscale eddies, a long-lived cyclonic eddy detected by the HEPTA algorithm in the period from August 30, 2019 to September 13, 2020 is considered. The $E_1$ eddy originated as a result of separation from the meandering Kuroshio Current (Fig. \ref{fig:2}a). The trajectory of the eddy center, reconstructed by the algorithm (Fig. \ref{fig:2}b) as of November 29, 2019, demonstrates a complex westward drift path. A key event in the eddy's life cycle was its division in mid-August (August 15, 2020), when a pair of stationary points formed in the peripheral zone~--- an elliptic $E_2$ and a hyperbolic $H_1$ (Fig. \ref{fig:2}c). The HEPTA algorithm recorded the splitting of the original elliptic point $E_1$ into two, $E_1$ and $E_2$, with the appearance of an intermediate hyperbolic point $H_1$, which corresponds to a "fork" type bifurcation. In the following months, the main eddy $E_1$ continued to drift north. The final stage of the eddy's evolution (Fig. \ref{fig:2}d) is associated with its merging with the main current. The algorithm recorded the collapse of the elliptic point $E_1$ and the hyperbolic point $H_2$ near the front, which corresponds to the absorption of the eddy by the background flow. On each panel of Fig. \ref{fig:2}, red triangles~--- anticyclonic eddies, green~--- cyclonic eddies, and crosses~--- hyperbolic points, detected by the HEPTA algorithm using AVISO altimetric data for the study region are additionally plotted.
	
	\section{Discussion}
	
	The proposed HEPTA algorithm uses bilinear and trilinear interpolation to determine stationary points in the vector velocity field. Compared to more complex methods, such as splines or high-order polynomial interpolation, bilinear and trilinear interpolation require significantly less computational cost and are easy to implement, while allowing some results to be obtained analytically. In most applied problems (oceanology, meteorology), measurement data contain noise and inaccuracies. The use of complex interpolation methods in such conditions will lead to the appearance of non-physical features. When predicting the trajectories of LCS in the ocean, the measurement error of velocity often exceeds the error of linear interpolation. It is important to note that this approach is not universal. It will not be applicable if the velocity field contains singularities or discontinuities, it is necessary to analyze high-frequency oscillations (for example, turbulence), as well as with a grid step that is too large, when nonlinear effects will dominate. In such cases, it is necessary to use more complex methods or increase the grid resolution.
	
	The proposed approach, where the position of stationary points was determined depending on the time parameter $\tau$, can be adapted to analyze other physical parameters. For example, applying a similar methodology to the spatial distribution of the velocity field with depth will allow reconstructing the vertical axis of the eddy~--- a key characteristic reflecting its stability, baroclinicity, and interaction with bottom currents.
	
	The HEPTA algorithm has significant potential for further development aimed at expanding analytical capabilities and increasing the information content of the results. In particular, it is planned to automate the calculation of kinematic characteristics, such as translational velocities, displacements, vertical scale, and a number of other parameters. A module will be added to determine the spatial dimensions of the areas of influence of stationary points (for example, for elliptic points~--- determining the closed contour of maximum rotational speed), which will allow for a more detailed analysis of the dynamics of complex systems. Further development of HEPTA also includes automatic tracking of eddy merging and division processes, which will significantly expand the possibilities for studying turbulence and other dynamic phenomena in the ocean.
	
	\section{Conclusion}
	The presented algorithm for tracking elliptic and hyperbolic stationary points in the altimetric velocity field has demonstrated its effectiveness in analyzing the dynamics of mesoscale eddies. By combining methods for determining the position, classification, and tracking of stationary points, the algorithm allows to quickly automate the process of identifying and monitoring Lagrangian coherent structures, providing a detailed study of their evolution in time. The results obtained using the HEPTA algorithm open up new opportunities for studying the processes of eddy formation, the interaction of eddies with large-scale currents, and their influence on the transfer of mass and energy in the ocean. Further development of the algorithm can be aimed at improving its computational efficiency, adapting to data with different resolutions, and integrating with other ocean models to obtain a more complete picture of the dynamics of oceanic processes.

	\section{Acknowledgements}
	
	The work of was supported by the Russian Science Foundation (project no. 23-17-00068) with the help of a high-performance computing cluster at the Pacific Oceanological Institute (State Task $\#124022100072-5$)

\end{document}